# Neuronal noise as a physical resource for human cognition

T.N. Palmer[1], M.O'Shea[2]
1. Department of Physics, University of Oxford
2. Centre for Computational Neuroscience and Robotics, University of Sussex

**A new class of energy-efficient digital microprocessor is being developed which is susceptible to thermal noise and consequently operates in probabilistic rather than conventional deterministic mode. Hybrid computing systems which combine probabilistic and deterministic processors can provide robust and efficient tools for computational problems that hitherto would be intractable by conventional deterministic algorithm. These developments suggest a revised perspective on the consequences of ion-channel noise in slender axons, often regarded as a hindrance to neuronal computations. It is proposed that the human brain is such an energy-efficient hybrid computational system whose remarkable characteristics emerge from constructive synergies between probabilistic and deterministic modes of operation. In particular, the capacity for intuition and creative problem solving appears to arise naturally from such a hybrid system. Bearing in mind that physical thermal noise is both pure and available at no cost, our proposal has implications for attempts to emulate the energy-efficient human brain on conventional energy-intensive deterministic supercomputers.**

Although the word `noise' carries negative connotations and conventional digital computers are designed to be as noise-immune as possible, there are situations where noise can be a constructive resource in algorithmic problem solving. Consider for example the Travelling Salesman's decision-making problem where the task is to visit all customers once and to return home having travelled the minimal possible distance. This problem is challenging because the number of candidate shortest routes increases exponentially with the number of customers. Heuristic (i.e. approximate) algorithms which incorporate some elements of randomness, in practice through pseudo-random chaos, can provide both efficient and robust techniques for finding solutions to such problems[1] (see Fig 1). One of the key reasons for the effectiveness of randomness is that it minimises the possibility of problem instances where a deterministic heuristic either proves misleading or takes an exceptionally long time to reach a solution[2]. In practice there may be good reasons to combine probabilistic and conventional deterministic algorithms into a single hybrid scheme, for example by using a stochastic search method (e.g. simulated annealing) to find the region of the search space likely to contain the global optimal solution, with a deterministic (e.g. gradient descent) algorithm then iterating to the optimal solution[3].

A programme to develop computers which operate in probabilistic rather than conventional bit-reproducible mode[4,5] is motivated by the fact that as the density of microprocessors in a computer continues to increase, and as individual transistors approach atomic scale, the overall power needed to ensure microprocessors operate deterministically is becoming larger, potentially unsustainably so. By relaxing so-called guardband constraints (which ensure a sufficient voltage across the transistors so that they are immune to internal or external noise), it is possible to design microprocessors that operate probabilistically rather than bit-reproducibly with a considerable reduction in energy consumption, thereby increasing the "Flops/Watt" metric of computational performance (see Fig 2). Combining a relatively large number of energy-efficient probabilistic processors with a relatively small number of energy-intensive deterministic processors can provide new, robust and computationally-efficient hybrid probabilistic\deterministic tools for solving complex (and otherwise intractable) computational problems[6,7].

As discussed below the ion channels or "protein transistors", which amplify electrical signals in neurons, are subject to thermal noise (ion channel noise) and therefore do not operate entirely deterministically[8]. In sufficiently slender axonal and dendritic arborisations such thermal noise can introduce a stochastic component to action potential generation[9,10]. Based on the developments in computer science outlined above, we propose here that this noise may have substantial beneficial consequences for the brain's computational performance and hence cognitive properties. Of course, unlike a digital computer, the brain evolved unconstrained by a designer's assumptions about how it ought to work – to operate purely deterministically, for example. So, if there were benefits to be gained by combining probabilistic and deterministic processes in the brain - for example in terms of increased energy efficiency and a smaller likelihood of "hanging" when making complex decisions - natural selection would have exploited them.

The ability of stochastic channel noise to corrupt the temporal pattern of a neuronal spike train, by random addition or deletion of impulses, increases as an inverse function of the diameter of axonal arborisations[10]. Experiments with axons of diameter greater than 1μ show that impulse generation is reliably deterministic. Such reliability however is costly because larger neurons, with their high speed information coding, are relatively energy inefficient. In neurons with the most slender axonal and dendritic arborisations (around 0.1μ), speed of information coding might be sacrificed for the benefit of increased energy efficiency. However, (as in transistors with reduced guardband voltage) energy savings may also be accompanied by stochastic corruption of the temporal pattern of impulse trains. Overall then, thermal noise affecting voltage sensitive ion channels is likely to decrease the reliability of impulse timing especially in fine axons when energy supply is restricted (Fig 3). From this discussion it would seem reasonable that neuron miniaturisation, and the accompanying benefit of higher packing density of computing elements, is limited by the trade-off between energy efficiency and the need to preserve information coding at an acceptable rate and reliability[11,12].

There have been proposals which have considered a constructive role for stochasticity in the brain, most notably through `Stochastic Resonance' or SR[13]. SR in threshold-determined sensory systems can for example enhance sensitivity to periodic fluctuations in the strength of sub-threshold signals in a noisy environment[14]. Here random noise resonates with the just sub-threshold periodic signal and increases the probability that repetitive bursts of spikes at the signal period will be generated. In this way, SR is a mechanism for making the electrical activity of neurons more reliably deterministic. This suggests a dichotomy in which noise is either beneficial for deterministic operation (SR) or is a nuisance (channel noise in very small neurons). By contrast, however, here we propose something more radical: that the inherent probabilistic character of signal corruption by channel noise actually contributes directly and positively to brain function.

If this proposal is correct, we might expect the neurons with the greatest susceptibility to signal corruption by channel noise to be engaged in solving the type of tasks that benefit most from hybrid computing, such as the classic combinatorial Travelling Salesman problem. Evidence that animals incapable of conscious deliberation can solve this problem comes from an unexpected direction. Experiments on bumblebees foraging on arrays of artificial flowers appear to optimise their flight distances and rearrange their flower visitation sequences dynamically as new sources of food are presented[15]. The bee brain is estimated to contain about one million neurons, a majority of which have branched axonal and dendritic compartments of sub 0.1µm dimensions. This places them well within the range where noise-induced corruption of spike timing can be expected. It is therefore plausible to hypothesise that a dynamic routing task, analogous to the travelling salesman problem, is being efficiently solved in part stochastically and presumably without conscious effort.

Could a model which combines stochasticity and determinism in some synergistic hybrid operation be relevant to the human brain? Indeed, could the relative degree of stochasticity/determinism in the brain be controlled by the effortful thinking usually referred to as `concentration'? Certainly functional imaging techniques show that the act of effortful thinking causes oxygenated blood to be diverted to specific local regions where increasing neural activity would otherwise outstrip energy supply. The fact that this must occur while the total blood supply to the whole brain remains substantially unaltered suggests that at a very broad-brush level, one might consider the brain operating between two cognitive modes of operation referred to here as Mode 1 and Mode 2. Mode 1 is an economical, relatively low-energy mode which maintains energy consumption across the many, small, efficient but slower neurons, susceptible to thermal noise. In Mode 2, available energy is focussed on a less efficient subset of neurons ensuring that they operate reliably, quickly and hence deterministically. It seems plausible to relate Modes 1 and 2 directly to Kahneman's[16] `fast/slow' System 1/2 paradigm of human thinking.

Below we offer a perspective on some of the possible consequences of hybrid probabilistic/deterministic operation in the brain, in particular for understanding human intuition and creativity. For example, it is a familiar experience that taking a break in concentration from some difficult problem, i.e. switching from Modes 2 and 1, can provide unexpected new angles on the problem which may ultimately lead to its solution. The mathematical physicist Roger Penrose[17] has documented a number of classical `eureka moments' when a scientist (himself included) was engaged in otherwise mundane activity, such as crossing the road or stepping onto a bus. If, in some non-deterministic way, a potential insight occurs when the brain is operating in Mode 1, it is straightforward to check using Mode 2 that this insight does indeed solve the problem.

To make these ideas more specific, consider the problem of how the human brain might go about proving the irrationality of $\sqrt{2}$. Firstly, although thermal in character, the ultimate origin of channel noise in slender axons is associated with the ubiquitous process of quantum decoherence on the molecular scale (related to the inherently non-algorithmic[18] collapse of the quantum wavefunction to a measurement eigenstate). The evolution of a dynamical system susceptible to such noise will not in general be closed (i.e. contained) within any finite subspace, as would be the case if stochasticity were represented by pseudo-noise generated from low-order chaos. Consistent with this, our intuitive appreciation that the set of integers $\{1,2,3...\}$ has no upper bound – often obtained in early childhood - can perhaps be considered a primitive product of Mode 1 operation of the brain. On the other hand, the notion that the decimal expansion of $\sqrt{2}$ similarly never ends is not itself something about which we have an intuitive understanding. Here we propose that the process of finding a proof of the irrationality of $\sqrt{2}$ requires a more sophisticated switching between probabilistic Mode 1 and deterministic Mode 2. For example, by looking up at the sky, one can imagine Mode 1 randomly introducing candidate ideas for Mode 2 to subsequently explore by deductive logic, perhaps based on the fractal geometry of clouds or the spherical geometry of the sun. After switching to Mode 2, these candidate proofs can be rejected as not readily leading to a proof or disproof of the problem at hand. However, a further small Mode 1 random iteration, which moves the brain's cognitive state from one which focuses on the potential relevance of geometry (spherical or fractal) to one which focuses on the potential relevance of number (rudimentary properties of odd and even numbers in particular), a further and final switch to Mode 2 could then reveal – as first discovered by the ancient Greeks - the logical proof of the irrationality of $\sqrt{2}$.

This bimodal probabilistic/deterministic hybrid hypothesis also provides a new perspective on the notion of free will – an essential element in defining the phenomenon of consciousness. The notion of free will poses a well-known dilemma summarised recently by quantum physicist Seth Lloyd[19] : `If determinism robs us of free will, then so does randomness'. That is to say, it is difficult to explain the perception of free will either from a purely deterministic (e.g. Newtonian) perspective, or from a purely stochastic (e.g. quantum mechanical) perspective. However, exploring synergistic hybrid relationships between stochasticity and determinism provides a natural and straightforward solution to this dilemma. Suppose for instance that every morning Bob must decide what shirt to wear. To keep matters simple, suppose Bob only wears polo shirts, stacked in a neat pile in his clothes drawer. The simplest and most frequently-made decision is to wear the shirt at the top of the pile. In this sense, a simple (deterministic) algorithm can determine with significant predictive skill the colour of the polo shirt Bob will wear. However, the algorithm will fail from time to time. For example, the colour of the polo shirt at the top of the pile may not be one he is especially keen on. In these situations, a Mode-2 analysis of which alternative shirt to pull out (What colour am I keen on, and why this colour rather than that?) may take considerable time and expend much energy for little real benefit. In such a situation, a much quicker and equally effective decision might instead be made in low-energy Mode 1 operation, where the decision-making neurons are susceptible to randomness. In these particular situations it would be natural to have the cognitive experience that `I could have done otherwise', the feeling of free will, since a counterfactual world which differs from the actual world only by the realisation of some particular random variable seems entirely plausible (though see[20]).

We conclude with some comments about links to artificial intelligence. Firstly, from a theoretical point of view, our hybrid probabilistic/deterministic computing system provides a novel way to understand the implications of Gödel's Theorem - that we humans can see the truth of mathematical propositions which cannot be proven by finite algorithm - for artificial intelligence. Lucas and Penrose[17] argue that because of Gödel's theorem, the human brain cannot be operating by finite algorithm and therefore cannot be replicated by a conventional digital computer, no matter how big. Penrose[12] has argued that coherent quantum entanglement effects must therefore be operating in the brain. However, there is little support for Penrose's thesis within the neuroscience community[21]; not least it is believed that quantum entanglement can play no significant role in the action of the brain because decoherence timescales $10^{-13}$-$10^{-20}$s in the warm noisy environment that is the brain would prevent isolated entanglements from lasting long enough to be relevant for dynamical neural timescales $10^{-3}$-$10^{-1}$s [22]. However, the Lucas/Penrose argument is readily explained in the hybrid probabilistic/deterministic proposal for the operation of the brain, precisely because the ultimate source of neuronal noise in the brain is non-algorithmic quantum decoherence at the molecular level.

From the perspective of the hybrid probabilistic/deterministic proposal, the purer the source of noise (ie the less it can be emulated by algorithm) and the less energy needed to access it, the more its effectiveness for complex problem solving. In stochastic search algorithms (such as simulated annealing discussed in Fig 1), the source of stochasticity is commonly based on deterministic low-order chaos. However, randomness associated with quantum decoherence is not only inherently purer than that from low-order chaos, as a physical resource it is ubiquitous and for all practical purposes can be (and indeed is) readily accessed with very low energy overhead. As a source of pure noise is available for free, it would not be surprising if evolution were to make use of this resource.

These remarks are relevant for attempts to emulate the brain on next-generation exascale computers[23]. Notwithstanding the fact that such computers are predicted to require in excess of 50MW to operate[24] and hence will need 6 orders of magnitude more energy than the brain itself needs, the results here suggest that thermal noise is an essential element in the operation of the brain, whose cognitive functions will therefore not be fully emulated on a purely deterministic machine. For some purposes pseudo-random noise may be good enough for emulation purposes, but experience with deterministic algorithms suggests that for complex optimisation problems, the use of low-dimensional chaos may, for some problem instances, cause the algorithm to effectively "hang". We encourage those seeking to emulate the brain to make explicit use of much purer physical thermal/quantum noise arising from the computer itself, and to consider ways in which such noise can provide a positive resource for complex problem solving, in the synergistic deterministic/stochastic sense described here.

Advanced Scientific Advisory Committee (ASCAC) Subcommittee. US Department of Energy. Office of Science. (2010)

# Figures

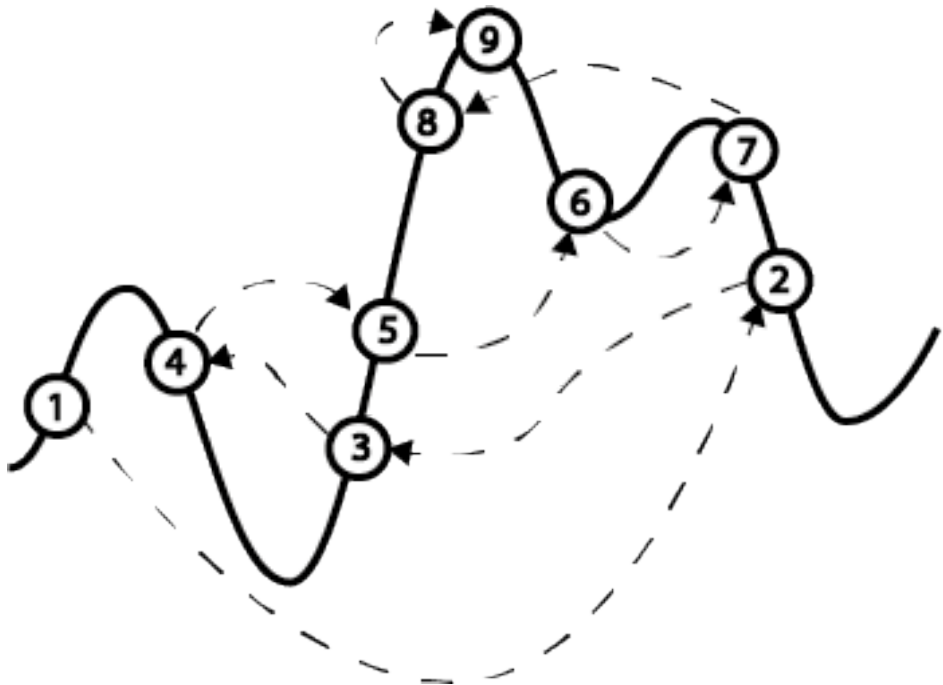

Figure 1

Simulated annealing is a probabilistic heuristic for the problem of finding the global optimum of a given function in a large search space. It is, for example, used to find solutions to the travelling salesman problem. At each step and using some source of noise, e.g. from a pseudo-random number generator, the heuristic decides probabilistically whether to move to some neighbouring state or remain at the current state. The probability of making the transition is determined by an acceptance function that depends on the "energies" of the two states and on a global time-varying "temperature" parameter which decreases to zero as the heuristic proceeds. Here a possible path 1➔2➔...➔9 is shown. In the early stages of the optimisation when the "temperature" is high (e.g. the transition from 2 to 3) it is possible for the new state to have lower energy than the old state. Such a transition is strongly penalised later in the optimisation procedure as the "temperature" drops to zero.

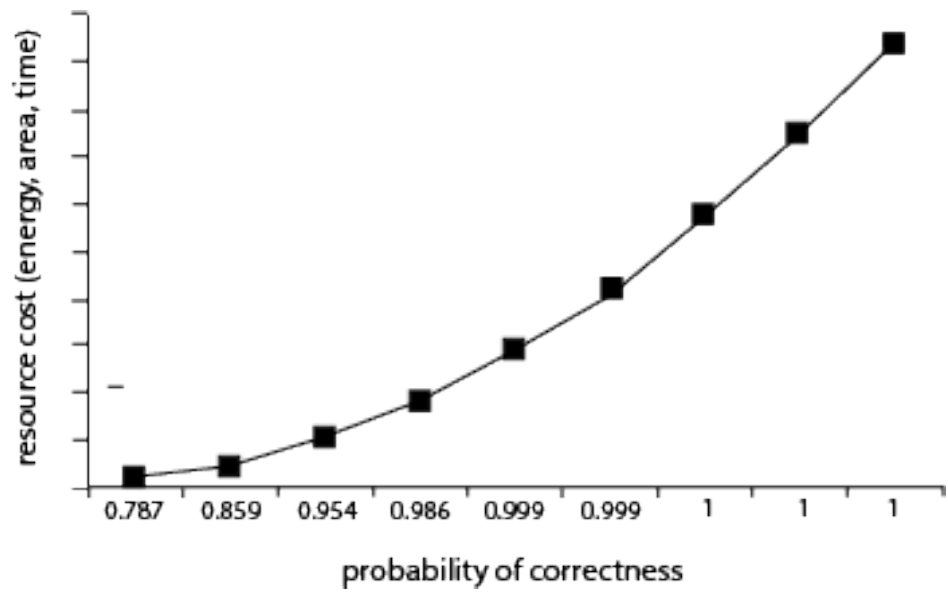

Figure 2

The estimated relationship between exactness and energy cost, for a class of probabilistic Complementary Metal-Oxide Semiconductor (CMOS) processors (from[4,5]).

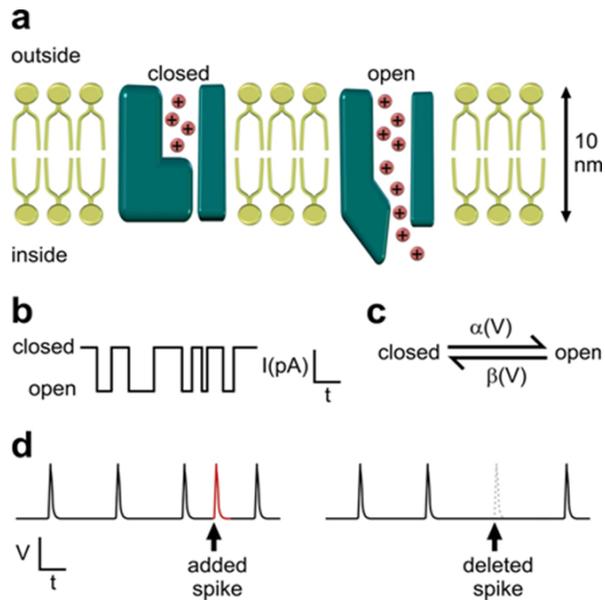

Figure 3

Specialised proteins form voltage-gated ion channels that span the 5nm thick neuronal lipid bilayer and regulate the flow of ionic current into and out of the neuron.  A simplified schematic of cation channels illustrates the closed and open state (a).  The transitions from close to open and open to closed is regulated by transmembrane voltage and underlies the deterministic generation of action potentials or spikes.  At a constant transmembrane voltage individual channels flicker stochastically between open and closed states (b) as governed probabilistically by the voltage dependent forward rate constant α and the backward rate constant β (c).  In axons below 1μm diameter single channel openings caused by channel noise can result in spike generation (added spike) or spike deletion (d).